\documentclass[onecolumn]{article}

\usepackage{graphicx}
\usepackage[top=2cm, bottom=2cm, left=2cm, right=2cm]{geometry}
\usepackage[textsize=footnotesize]{todonotes} 
\usepackage{upgreek}
\usepackage{authblk}
\usepackage[font={small},labelfont=bf]{caption}
\usepackage{fixltx2e}
\usepackage{amsmath}
\usepackage{textcomp}
\usepackage{verbatim}
\usepackage{lineno}

\usepackage[square,super,comma,numbers,sort&compress]{natbib}
\bibpunct{}{}{,}{s}{}{\textsuperscript{,}}

\title{\textbf\noindent{Magnetic topological lithography: Gateway to the artificial spin ice manifold}}

\renewcommand{\deg}{$^{\circ}$ }

\begin{document}
\author{J. C. Gartside\textsuperscript{1}\thanks{j.carter-gartside13@imperial.ac.uk}}
\author{D. M. Arroo\textsuperscript{1}}
\author{D. M. Burn\textsuperscript{2}}
\author{V. L. Bemmer\textsuperscript{3}}
\author{A. Moskalenko\textsuperscript{1}}
\author{L. F. Cohen\textsuperscript{1}}
\author{\\ W. R. Branford\textsuperscript{1}}
\affil{Blackett Laboratory, Imperial College London\textsuperscript{1} \\ Diamond Light Source, Didcot\textsuperscript{2} \\ Department of Materials, Imperial College London\textsuperscript{3}}

\renewcommand\Authands{ and }

\maketitle


\begin{abstract}

\noindent Nanomagnetic arrays are widespread in data storage and processing. As current technologies approach fundamental limits on size and thermal stability, extracting additional functionality from arrays is crucial to advancing technological progress. One design exploiting the enhanced magnetic interactions in dense arrays is the geometrically-frustrated metamaterial `artificial spin ice' (ASI). Frustrated systems offer vast untapped potential arising from their unique microstate landscapes, presenting intriguing opportunities from reconfigurable logic to magnonic devices or hardware neural networks. However, progress in such systems is impeded by the inability to access more than a fraction of the total microstate space. Here, we present a powerful surface-probe lithography technique, magnetic topological lithography, providing access to all possible microstates in ASI and related nanomagnetic arrays. We demonstrate the creation of two previously elusive states; the spin-crystal ground state of dipolar kagome ASI and high-energy, low-entropy `monopole-chain' states exhibiting negative effective temperatures.

\end{abstract}

\noindent Artificial spin ices (ASI) are magnetic metamaterials comprised of strongly-interacting nanomagnets, each acting as a single `macrospin', in geometrically frustrated arrays. The macrospins obey magnetic equivalents of Pauling's ice-rules\cite{pauling1935structure}, arranging themselves to minimise net magnetic charge $q_m$ at each vertex. ASI\cite{Nat_439_303} has provided vital physical insight on diverse topics from magnetic monopole-like excitations\cite{Ladak2010,N.Phys_7_68} to fundamental thermodynamics\cite{Nisoli2007,Qi2008,Morgan2010,Lammert2010,Branford2012,Farhan2013}. Key to its promise is the enormous number of unique magnetic charge configurations (or microstates) and the possibility to directly image the magnetic structure. A kagome (honeycomb) ASI array containing N vertices possesses $\sim\frac{3}{\sqrt{2}}^\text{N}$ microstates satisfying the ice-rules, yielding vast microstate landscapes. The ice-rule obeying states all possess near-equivalent energies leading to massive degeneracy, inviting deep thermodynamic studies\cite{Farhan2013} as well as device-focused applications exploiting the microstate landscape to store information\cite{budrikis2012network,Heyderman2013} or function as a hardware neural-network\cite{Nisoli2007,Branford2012,NJP_13_023023} or reconfigurable magnonic crystal\cite{grundler2015reconfigurable,krawczyk2014review}. However, the full potential of ASI is unrealised as conventional state-preparation techniques access only a tiny fraction of the total microstate space. Consider a six-macrospin hexagon of kagome ASI, possessing $2^6 = 64$ metastable magnetic configurations (local minima in the potential energy surface). Typically a global magnetic field initialises state $A$ and then drives magnetic reversal to state $B$. As each bar switches only once, six additional states are sampled and so 57 states are never accessed. Energy barriers between states vary largely so few low-energy pathways from $A$ to $B$ exist. Most states do not lie on any of these pathways, rendering them inaccessible via global magnetic field protocols. In a typical $10^3$ macrospin array just $ 10^{-296}\%$ of possible states are sampled in a global-field driven magnetic reversal. Thermal activation and subsequent annealing of ASI can randomly sample low-energy microstates inaccessible using global fields\cite{Zhang2013,Farhan2013,farhan2013exploring,farhan2014thermally,kapaklis2014thermal}. However, the exact final state cannot be specified and to-date it has not been possible to explore the network of microstates sufficiently to prepare a system in the dipolar kagome ASI ground state\cite{farhan2014thermally}. The limited state-writing functionality has impeded progress in ASI, therefore developing a protocol for the precise control of individual macrospins is crucial to revealing  the entire  ASI manifold and the rich physics within.

Scanning-probe lithography (SPL) offers an ever-broadening suite of techniques for exquisite nanoscale  patterning\cite{garcia2014advanced}, including topographical modification\cite{eigler1990positioning,van1989direct}, molecular  synthesis\cite{pavlivcek2017synthesis} and recently nanomagnetic writing. Lithographic modification of magnetic anisotropy in thin-films via a heated  tip and global H field\cite{albisetti2016nanopatterning}, magnetic-force microscope (MFM) tip-mediated domain wall (DW) injection\cite{gartside2016novel} and a combined MFM-tip and global-field protocol for selective remagnetisation of nanowires\cite{wang2016rewritable} represent key advances in the field, forging paths for new work in fundamental and device physics. Here, we present `magnetic topological lithography' (MTL) - a novel, powerful magnetic SPL technique for directly injecting topological defects\cite{mermin-topology} in magnetic nanowires. These defects propagate through structures as DWs, achieving controlled and reversible writing of magnetisation states in the process with no global fields required. Working in ambient conditions, MTL reveals the entire microstate landscape of ASI and related nanopatterned magnetic systems. We demonstrate several previously unobserved states including `monopole chains', high-energy low-entropy ASI configurations inaccessible by conventional means and exhibiting negative effective temperatures\cite{nisoli2010effective}, and the spin-crystal ground state of kagome ASI\cite{Zhang2013,farhan2014thermally,anghinolfi2015thermodynamic}.

\section*{Working principle of MTL}

\begin{figure}[tbp]
	\centering
	\includegraphics[width=12cm]{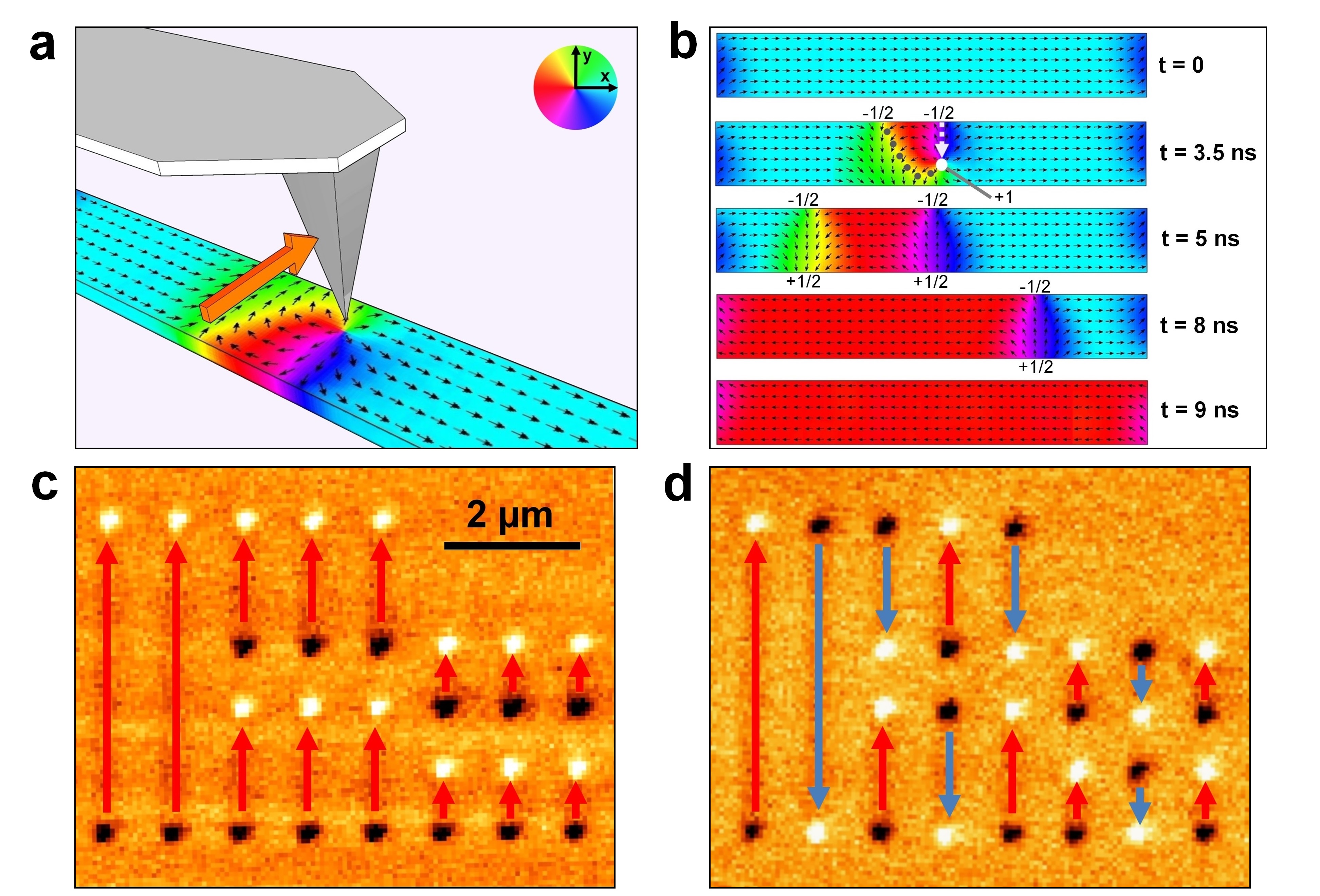}
\caption{
\textbf{The MTL writing process.} \textbf{a)} Schematic of MTL operation. An MFM-tip induced topological vortex defect is moved through a nanostructure spin texture by the motion of the tip. Colour wheel represents magnetisation direction in panels a) and b), orange arrow denotes $v_{tip}$.\\
\textbf{b)} Micromagnetic simulation of the time-evolution of the MTL write-function used to reverse the magnetisation of a $500 \times 75 \times 10$ nm nanowire. The MFM tip and its direction of motion are represented by the white circle and dashed arrow respectively. Topological defects are labelled with their winding numbers. A video of this process is available in the supplementary materials. \\
\textbf{c, d)} MFM images of 10 $\times$ 100 $\mathrm{nm}^2$ cross-section NiFe nanowires of 1, 2 and 4.8 $\upmu$m length, (c) after initialisation with global field and (d) after MTL writing of selected wires. Light and dark contrast indicate positive and negative magnetic charge respectively. Overlaid arrows show magnetisation direction as a guide to the eye, red arrows indicate unwritten wires, blue arrows show MTL-written wires.
}
\label{fig1}
\end{figure}

MTL is depicted schematically in fig. \ref{fig1} a) with a corresponding time evolutionshown in fig. \ref{fig1} b) (details of the OOMMF\cite{oommf} micromagnetic simulations given in the supplementary materials). We demonstrate MTL experimentally first by writing states in NiFe nanowires. Figure \ref{fig1} c) shows MFM images of nanowires after initialisation in global field $H_{sat}$, 1 d) shows MFM after selective MTL reversal. MTL consists of passing a commercially available high-moment MFM tip over a nanostructure at a tip-sample separation of $r \approx 5-10$ nm with the tip moving perpendicular to the long-axis of the structure. At these small tip-sample separations, the tip field $H_{tip}$ may be treated using a point-probe approximation\cite{Phys.Lett.A_137_475,JAP_86_3410} as having a monopole-like form $H_{tip} = \frac{\mu_0}{4 \pi r^2}q_{tip}$ where $q_{tip}$ is the magnetic charge of the tip. The nanowire begins the process in a collinear ferromagnetic state (t = 0 in fig. \ref{fig1} b). The tip then approaches the structure and at sufficient values of $H_{tip}$ the spins in the wire rearrange themselves to lie along the radial tip field, lowering their Zeeman energy. Around the apex of the tip (t = 3.5 ns, represented by the white circle) these realigned spins form a vortex configuration due to the locally divergent shape of $H_{tip}$. In an otherwise smooth spin system, a vortex is a topological defect with an associated winding number\cite{PRL_95_197204} of +1. The net winding number of a ferromagnetic spin system is rigorously conserved and as such, the formation of the vortex is balanced by creating a pair of half-vortex topological defects with winding number -1/2. Defects with fractional winding number are forbidden in the bulk so the half-vortices remain bound to the edge of the structure\cite{PRL_95_197204}. The +1 vortex remains under the tip as it moves through the nanostructure, maintaining a connection to the edge-bound half-vortices via two continous chains of reversed spins to avoid a magnetisation discontinuity. The chains are highlighted in the t = 3.5 ns panel by the dotted grey line (left of tip) and the dashed white line (in line with tip motion). The tip effect is asymmetric: On one side (right of tip in fig. \ref{fig1} b) the component of $H_{tip}$ along the wire-length is parallel to the wire magnetisation, so any disturbance to the micromagnetic structure is minimal. On the other side $H_{tip}$ is anti-parallel to the magnetisation and drives local magnetic reversal. This forces the left-hand chain into an elongated curve around a nascent left-magnetised domain created between the -1/2 defects (t = 3.5 ns) in the wake of the tip. As the tip finishes crossing the structure, the vortex is brought into contact with the structure's lower edge. Whole integer defects may only exist in the bulk, so the vortex decomposes into a pair of edge-bound +1/2 defects to conserve the net winding number\cite{PRL_95_197204}. The chains of reversed spins now each connect a +1/2 topological defect on the bottom edge of the structure to a corresponding -1/2 defect on the top edge (t = 5 ns), each chain-defect pair forming an independent 180\deg DW\cite{PRL_95_197204}. The distorted shape of the left-hand chain of spins has an associated micromagnetic exchange and demagnetisation energy penalty relative to the straight, shorter right-hand chain. In the absence of the tip the left-hand chain straightens out, adopting a lower energy conformation and gaining momentum in the process. The DW accelerates towards the left-end of the wire, aided by a magnetostatic attraction between the DW magnetic charge and the magnetic charge associated with the wire-end. On reaching the left-end of the structure, the DW is free to unwind into a collinear spin state (t = 8 ns), lowering the total system energy. The remaining right-hand DW also experiences a magnetostatic attraction to the structure ends. Providing the tip crosses the structure slightly to the right of centre, the net force on the remaining DW is directed to the right. This pulls the DW into the right-hand end of the structure where it unwinds on contact, leaving the wire DW-free, magnetised anti-parallel to its initial state. The MTL writing process is now complete. Whether the right or left-hand DW forms in a high-energy state is decided by a combination of the nanostructure's initial magnetisation and the magnetisation of the MFM tip, allowing the user to tailor the writing dynamics simply by remagnetising the tip. The combination depicted in fig. \ref{fig1} b) arises from a right-magnetised nanostructure and MFM tip magnetised into the page. Reversing either magnetisation will result instead in the right-hand DW forming in a high-energy state and accelerating once the tip passes. Reversing both magnetisations returns to a high-energy left-hand DW. Simulated videos of this process are provided in the supplementary materials. Crucially, the MTL process relies solely on the local field of the tip with no additional global $H$ field required, enhancing utility and flexibility relative to existing techniques\cite{wang2016rewritable}. 

\begin{figure}[tbp]
	\centering
	\includegraphics[width=15cm]{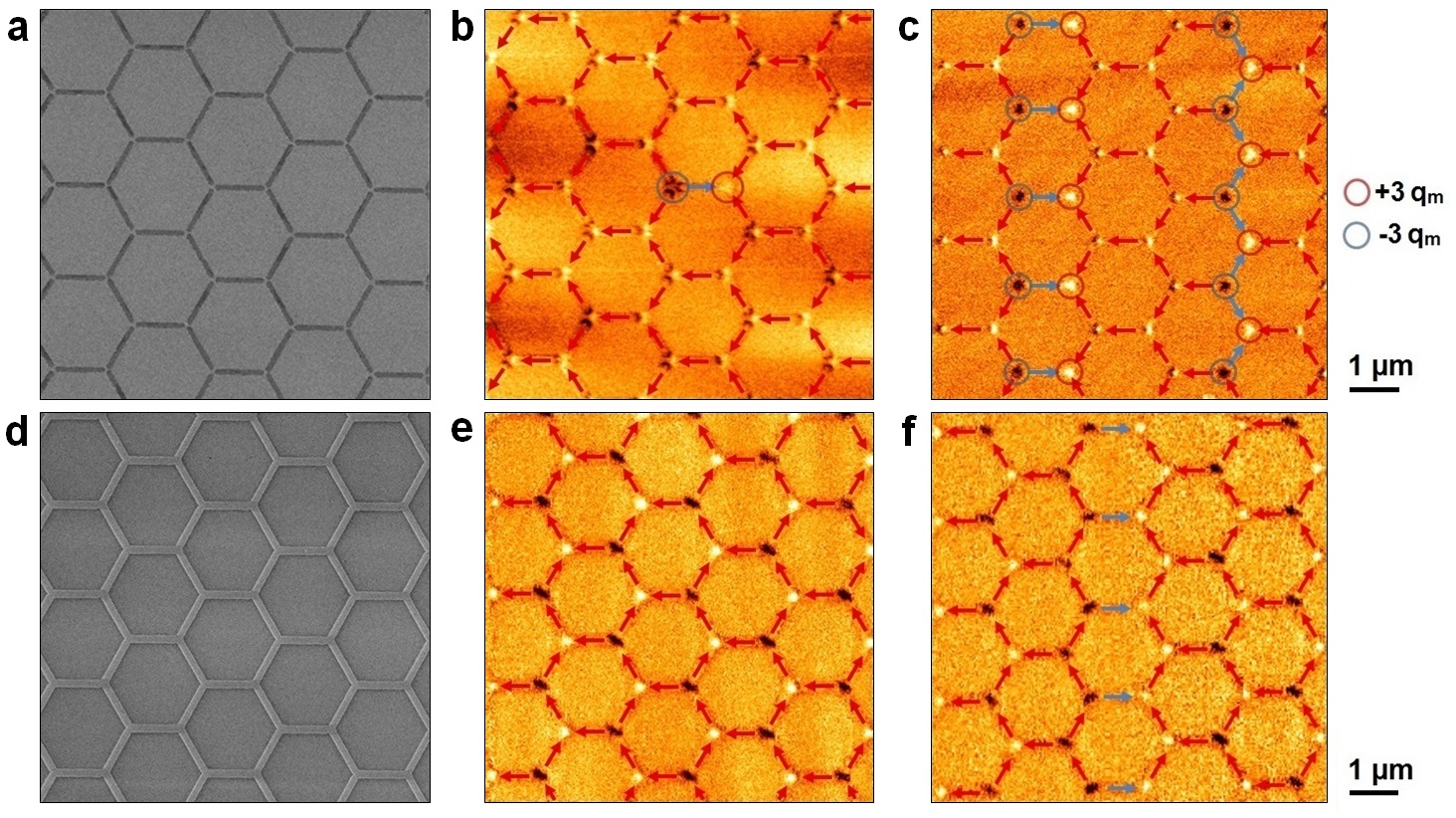}
\caption{\textbf{Realisation of MTL read/write functionality in ASI.} SEM (a,d) and MFM (b,c,e,f) images demonstrating MTL written states in various ASI arrays. NiFe nanowires are 1 $\upmu$m long, 10 nm thick. a,b,c) wires are 75 nm wide, disconnected at vertices with 30 nm gaps. d,e,f) are 100 nm wide, connected at vertices. MFM images are overlaid with arrows indicating magnetisation direction of each macrospin. Prior to imaging and writing, structures were saturated by an external field $H_{Sat}$. Red arrows denote unwritten nanowires still magnetised along $H_{sat}$, blue arrows show lithographically remagnetised wires.\\
\textbf{b)} MFM image showing a single tip-reversed nanowire, creating a pair of $\pm 3~q_m$ monopole-defects.
\textbf{c)} MFM image of lithographically written `ladder' and zigzag `monopole-chain' states. System was re-magnetised along $H_{Sat}$ between images (b) and (c).\\
\textbf{e)} MFM `before' image showing all wires aligned to $H_{sat}$. 
\textbf{f)} MFM `after' image of (e) showing a column of MTL-reversed nanowires.\\
}
\label{mfm}
\end{figure}

\section*{MTL state writing in ASI}

To examine how the MTL state-writing functionality performs in an ASI system, 1 $\upmu$m $\times~75~\times~10~\textrm{nm}^2$ nanowires were arranged into kagome ASI arrays with wires disconnected at vertices. Vertex separations of 30 nm were used, defined from the wire end to the vertex centre. Arrays were confirmed to be in a strongly-interacting ASI regime via MFM imaging of their as-grown states, with only ice-rule obeying vertices observed.
Structures were initially saturated along a globally-applied field $H_{sat}$. MTL was then performed in zero-field to selectively reverse the magnetisation of specific macrospins until the desired microstate was reached. Figure \ref{mfm} shows scanning-electron microscope (SEM) images of the ASI arrays (left-hand column) along with MFM images of initialised and MTL-written states. Figure \ref{mfm} b) shows a single MTL-reversed macrospin against a uniform $H_{sat}$ aligned background, creating a pair of oppositely charged $\pm 3~q_m$ monopole-defect vertices without disturbing surrounding macrospins. This demonstrates both the spatial accuracy of MTL, allowing control of each individual macrospin, and its ability to overcome the significant energy-barriers required to create ice-rule breaking states in strongly-interacting systems, thus granting access to the full ASI manifold. Figure \ref{mfm} c) shows a more complex MTL-written state containing long strings of adjacent ice-rule violating vertices, previously unobserved `monopole-chains'. Demonstrated here are both `ladder' (left) and zigzag (right) monopole-chains, high-energy states (due to the large number of excited $\pm 3~q_m$ vertices) with low entropy (due to the relatively few ways to arrange such configurations on an ASI array). This combination of high energy with low entropy results in a state exceedingly difficult to realise via conventional means and also fulfills the criteria for exhibiting `negative temperature' - predicted to exist in ASI\cite{nisoli2010effective} and discussed further below. Such states highlight the novel and unexplored regions of the ASI manifold now accessible using MTL. Even in these high-energy states the system retains its written configuration with high stability, undisturbed by the MFM imaging process and maintaining microstate fidelity over several months.

So far, only disconnected nanowires have been considered. To test the flexibility of MTL in varied architectures, ASI arrays were fabricated with nanowires connected at vertices. Figure \ref{mfm} d) is an SEM image of 1 $\upmu$m $\times~150~\times~10~\textrm{nm}^2$ nanowires in a connected kagome ASI array. Figure \ref{mfm} e) shows MFM imaging of the same array with macrospins saturated along $H_{sat}$. MTL was then used to reverse a central column of macrospins, with fig. \ref{mfm} f) showing the successfully-written resultant state. The MTL writing process is seen to retain its functionality in systems comprising connected nanowires, opening up microstate writing in a huge range of potential systems.

\section*{MTL realisation of the kagome ASI ground state}

\begin{figure}[tbp]
	\centering
	\includegraphics[width=15cm]{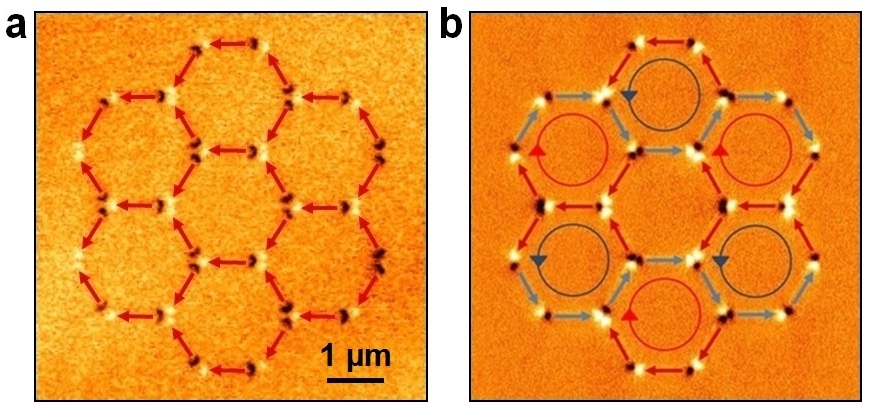}
\caption{\textbf{MTL-realised access to the kagome ASI ground state} 
\textbf{a)} MFM image of 1 $\upmu$m $\times$ 75 nm $\times$ 10 nm NiFe nanowires in an ASI 'rosette' with 30 nm vertex gaps. All wires are aligned to $H_{Sat}$. \\
\textbf{b)} MFM image of the chiral spin-crystal ground state of kagome ASI, accessed via magnetic topological lithography. Reversed wires are highlighted by blue arrows. Circular arrows inside hexagons highlight the alternating-chirality magnetisation loops characterising the ground state.}
\label{mfm2}
\end{figure}

One ASI microstate which has received significant attention while eluding observation is the spin-crystal ground state of dipolar kagome ASI. The ground state of the frustrated kagome lattice has been of interest in both natural and artificial systems, with the ground state of natural systems consisting of point-spins only recently reported after years of investigation\cite{liao2017gapless}. Point-spin systems interact via nearest-neighbour exchange interactions, hence the difficulty in identifying a long-range ordered ground state. However, when considering dipoles of finite length (i.e. artificial macrospins), further neighbour interactions (multipolar terms) break the degeneracy of ice-rule states as long-range magnetic charge and spin order are imposed\cite{moller2009magnetic,chern2011two,rougemaille2011artificial}. This leads to a well-defined ground state in ASI, existing when both charge and spin order are present throughout the entire system. A quasi-ground state in a partial `building block' kagome ASI system comprising just 1-3 hexagons has been demonstrated\cite{arnalds2012thermalized}, but the smallest system supporting a true ground state unit-cell which can tile infinitely over the kagome lattice is the 30 macrospin `rosette' shown in figure \ref{mfm2}. Attempts have been made to thermally activate then anneal into the ground state on a rosette-sized system\cite{farhan2014thermally}, but the near-degenerate energies of all ice-rule obeying microstates and the low entropy of the ground state (2 configurations in $2^{30}$) have rendered efforts ineffectual. A simple and convenient means to reliably access the spin-crystal ground state will allow closer examination of the divergence between true and artificial ice-like systems, a major point of interest in the field. As such, demonstrating the first experimental realisation of the spin-crystal kagome ground state is a perfect opportunity for testing the capabilities of MTL.

Figure \ref{mfm2} a) shows an MFM image of a kagome ASI rosette comprising 1 $\upmu$m $\times~75~\times~10~\textrm{nm}^2$ nanowires with all macrospins aligned along $H_{sat}$. MTL was then performed in zero-field to lithographically remagnetise 14 of the 30 macrospins, with the resultant state shown in fig. \ref{mfm2} b). The macrospins in each of the outer hexagons now lie in closed continous spin-loops, with chirality alternating between adjacent hexagons (highlighted by circular arrows within hexagons). This configuration not only satisfies the ice-rules but also exhibits magnetic charge and spin ordering, with the observed charge and spin configurations exactly matching those previously reported in theoretical schematics of the spin-crystal ground state\cite{moller2009magnetic,chern2011two,chern2012magnetic,farhan2014thermally,anghinolfi2015thermodynamic}. As such, we conclude that the MTL-written state presented here represents an experimentally-realised kagome ground state. That states eluding observation up to this point, despite direct efforts to prepare them, may now be written to order using readily-available equipment is an important threshold development for the fields of ASI and SPL.

\section*{Negative effective temperature states in ASI}

\begin{figure}[tbp]
	\centering
	\includegraphics[width=14cm]{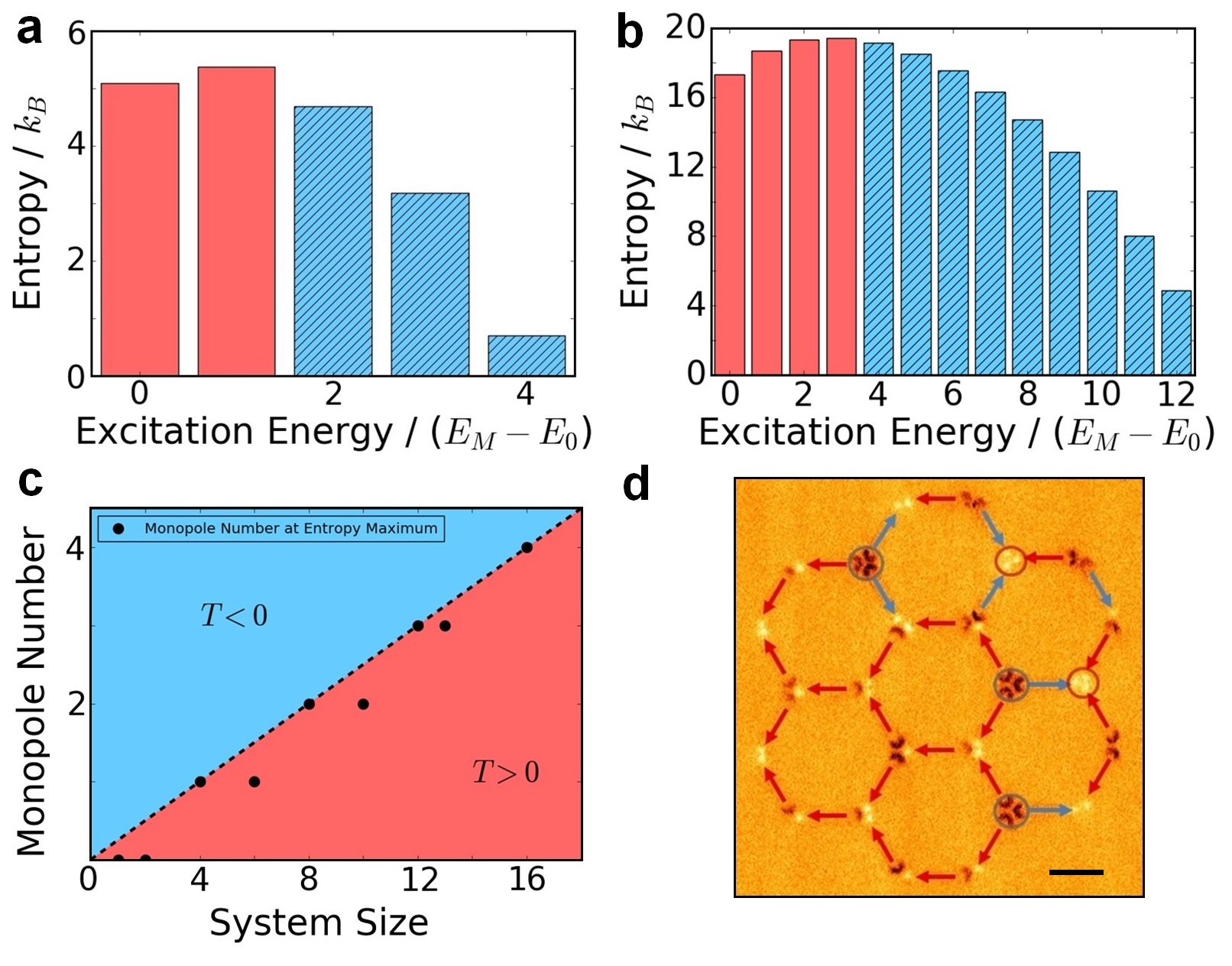}
\caption{\textbf{Negative effective temperature in ASI}\\
Entropy vs. excitation energy histograms for \textbf{a)} a four-vertex and \textbf{b)} a twelve-vertex kagome ASI system. Excitation energy normalised by $E_M - E_0$ corresponds to number of monopole vertices present. Red bars denote postive temperature states, shaded blue bars indicate negative effective temperatures.\\
\textbf{c)} Effective temperature phase diagram for one-to-eighteen vertex kagome ASI systems. Negative effective temperatures arise when more than one-quarter of total vertices are in monopole-defect configuration. \\
\textbf{d)} MFM image of a twelve-vertex (counting three-macrospin vertices only) kagome ASI rosette. Five monopole-defect vertices are observed, indicating a negative effective temperature state. Scale bar indicates 1 $\upmu$m.
}
\label{neg-temp}
\end{figure}

As well as providing reliable access to low-energy states, MTL also significantly expands the accessible ASI microstate space by allowing preparation of high-energy, low-entropy microstates such as the monopole-chains discussed above.
In general, the thermal anneals used to prepare ASI states work by keeping ASI systems in thermodynamic equilibrium with an (effective, in the case of field protocols) heat bath. As the temperature of the heat bath is reduced, the ASI settles into a low-energy state with relatively few ice-rule defects. Rotating-field protocols may be thought of in a similar way by assigning them an effective temperature based on the field step-size and treating them as a heat bath\cite{nisoli2010effective}. In contrast to these methods, MTL can reliably prepare ASI in states that are inherently out-of-equilibrium. For example, it is vanishingly improbable that a microstate with 19 monopoles on a system of 40 honeycomb vertices as shown in figure \ref{mfm} c) would occur through any sort of equilibration process with a heat bath of finite temperature. 
To formalise this idea, it is useful to invoke the concept of an effective spin temperature for ASI as introduced by Nisoli \textit{et al}\cite{nisoli2010effective}. In statistical mechanics, the temperature of a system is formally defined as a measure of how the energy of that system varies with disorder $T = \frac{\partial U}{\partial S}$. Taking into account only nearest-neighbour interactions (which dominate ASI energetics), the energy of an ASI system is determined by the number of monopoles (ice-rule violations) in the system, with each monopole raising the total energy by the difference in energy between a monopole and ice-rule vertex, $\Delta E = E_M - E_0$. The total energy of the system is thus given by $E = E_{ice} + n \Delta E$, where $E_{ice}$ is the system energy when all vertices obey ice-rules and $n$ is the number of monopoles. Since the fully ice-rule compliant state is highly degenerate but the fully excited state (where every vertex hosts a monopole) has just two possible configurations (spin-degeneracy), there must exist a range of microstates with negative effective temperatures\cite{purcell1951nuclear,abraham2017physics}.
Such states are precisely those difficult to access via thermal and rotating field protocols, but trivial to realise via MTL. 

By directly enumerating the number of monopoles (and therefore system energy) for all microstates in small kagome ASI systems and plotting monopole number versus entropy (figure \ref{neg-temp} a-c), the sign of the effective temperature may be clearly visualised from the gradient $\frac{\partial U}{\partial S}$. Figures \ref{neg-temp} a) and b) show this relation between entropy and energy/monopole-number in four and twelve-vertex systems respectively, with a clear change in gradient and associated transition to negative temperature occurring in both when over a quarter of vertices host monopole defects. The effective temperatures of all microstates in 1-18 vertex systems were calculated, with results shown as a phase diagram in fig. \ref{neg-temp} c). For all system-sizes negative temperatures were seen to correspond to microstates in which over a quarter of total vertices were monopole-defects. The five-monopole microstate on a 12-vertex honeycomb rosette shown in figure \ref{neg-temp} d) thus corresponds to a negative temperature state directly written by MTL, as does the monopole-chain state shown in fig. \ref{mfm} c). By comparison, the rosette system in fig. \ref{neg-temp} d) would have to be heated to almost $10^5$ K and cooled near-instantaneously to have a $0.1 \%$ chance of leaving the system in a five-monopole state. ASI enables direct room-temperature access to the complex and exotic physics of frustrated magnetism, with system energetics and time-scales easily tunable via fabrication. Adding to prior observations of otherwise-elusive ice-rule behaviour\cite{Qi2008} and monopole-defects\cite{Ladak2010}, MTL now introduces ASI as a model-system in which to study the dynamics and evolution of negative temperature states.

\section*{Conclusions}

In this article we have presented an outline and demonstration of MTL, a powerful and simple-to-use lithographic technique broadening the field of magnetic-SPL and allowing unprecedented access to novel and exotic states in ASI and related nanomagnetic systems. Reconfigurable magnetic metamaterials and the associated technological benefits are now a realistic and practical prospect, setting the scene for a period of rapid development as long-proposed designs including hardware neural-networks and reconfigurable magnonic crystals may now be experimentally realised. The spin-crystal ground state and negative-temperature monopole-chain states written in this work are evidence that all possible magnetic configurations are now accessible, both previously studied yet hard-to-access states through to states exhibiting exciting new physics, considered unattainable until now.

While the ultimate throughput of any SPL technique is limited by the mechanical nature of a moving tip, the magnetic charge supplied here by an MFM tip may come from a number of sources. A DW in a nanowire of the dimensions discussed here has $q_m \approx 10^{-9} Am$, within the limits for achieving MTL. One can envisage a system comprising a three-dimensional network of nanowires whereby current-controlled DWs replace the MFM tip, greatly enhancing flexibility, throughput and integration with existing technologies.

\section*{Author contributions}

JCG with the assistance of DMA, DMB and WRB conceived the experiment, JCG fabricated the samples and performed the experimental measurements, JCG, VB and AM developed the practical lithographic capabilities on SPM systems, JCG, DMA and DMB performed micromagnetic simulations, JCG performed the experimental measurements. JCG, DMA, WRB and LFC contributed to the manuscript. All authors contributed to discussions informing the research.

\section*{Acknowledgements}

This work was supported by the Engineering and Physical Sciences Research Council [grant number EP/G004765/1] and the Leverhulme Trust [grant number RPG 2012-692] to WRB and supported by the Engineering and Physical Sciences Research Council [grant number EP/J014699/1] to LFC.

\section*{Data statement}

Data requests should be addressed to dataenquiryexss@imperial.ac.uk.

\section*{Competing interests}

The authors declare no competing financial interests.

\bibliography{library}

\end{document}


\author{J. C. Gartside\textsuperscript{1}\thanks{j.carter-gartside13@imperial.ac.uk}}
\author{D. M. Arroo\textsuperscript{1}}
\author{D. M. Burn\textsuperscript{2}}
\author{V. L. Bemmer\textsuperscript{3}}
\author{A. Moskalenko\textsuperscript{1}}
\author{L. F. Cohen\textsuperscript{1}}
\author{\\ W. R. Branford\textsuperscript{1}}
\affil{Blackett Laboratory, Imperial College London\textsuperscript{1} \\ Diamond Light Source, Didcot\textsuperscript{2} \\ Department of Materials, Imperial College London\textsuperscript{3}}

\renewcommand\Authands{ and }

\maketitle

\section*{Supplementary information}

\section*{Micromagnetic dynamics of the MTL writing process}

\begin{figure}[tbp]
	\centering
	\includegraphics[width=16cm]{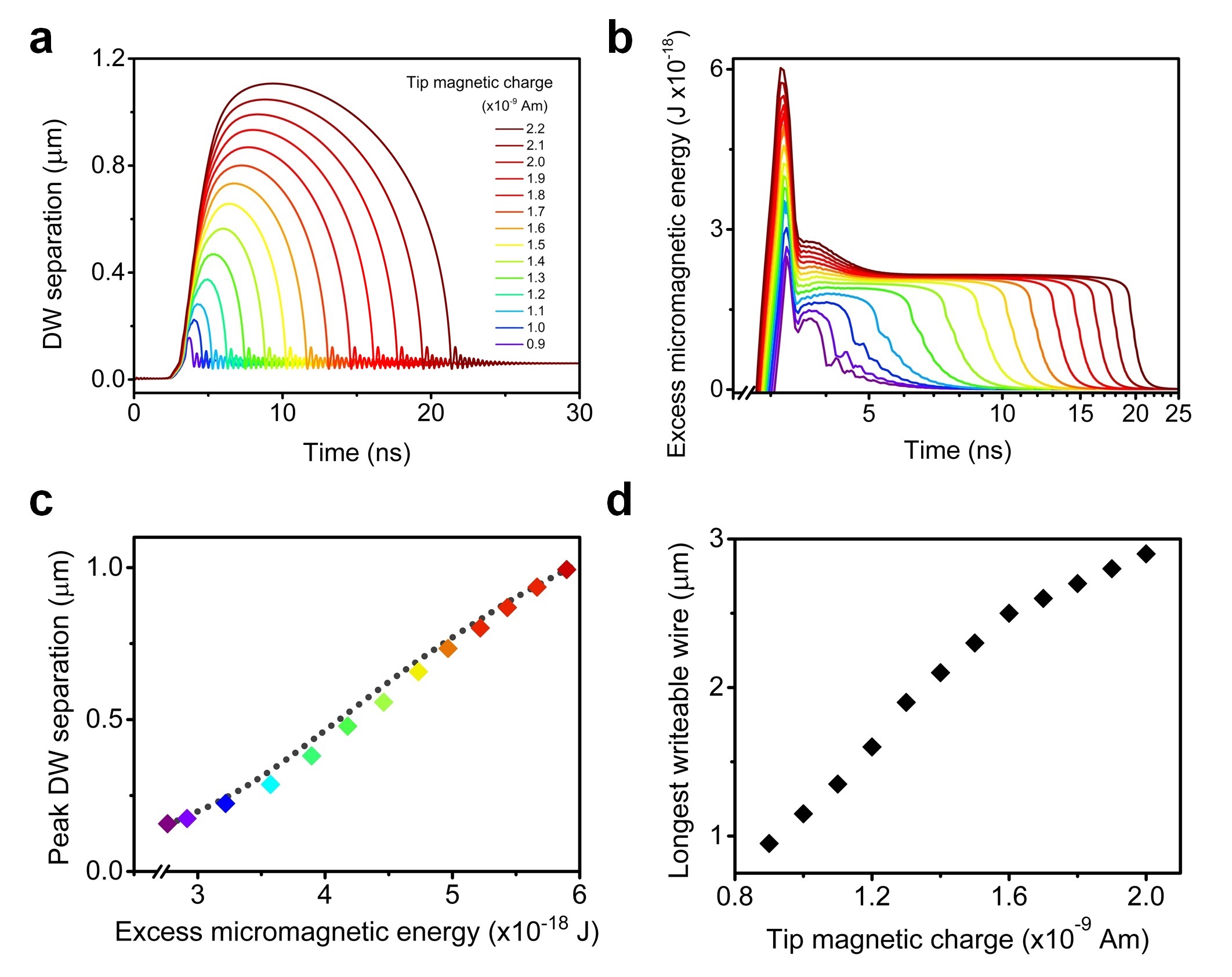}
\caption{\textbf{Dynamics of the MTL writing process.}\\
\textbf{a)} Time-dependent separation between domain walls injected in an infinite 75 nm wide nanowire by varying strength MFM tips. Tip reaches the wire at $t = 2.5~\mathrm{ns}$.\\
\textbf{b)} Excess micromagnetic energy versus time with colours corresponding to the same range of tip strengths as a).
\textbf{c)} Peak inter-DW separation for different strength MFM tips as a function of peak excess micromagnetic energy. Coloured diamonds are data taken from the OOMMF simulations shown in panel a) with colours corresponding to the same tip strengths. Grey dotted line shows data given by a simple semi-analytical kinematic model.\\
\textbf{d)} Longest simulated MTL-writeable nanowire versus MFM tip magnetic charge strength $q_{tip}$.
}
\label{sim}
\end{figure}

To gain further insight into the dynamics of the writing process, micromagnetic simulations were performed using OOMMF. To isolate the intrinsic MFM tip specific dynamics from the effects of the magnetic charges at the nanowire ends, a semi-infinite wire of 10 $\upmu$m length was simulated with plates of fixed magnetisation employed at either end to effectively remove the magnetic charges\cite{IEEE.Trans.Magn_33_4167}. A magnetic charge $q_{tip}$ and associated monopolar $H$ field representing the MFM tip were passed over a $75 \times~10 \mathrm{nm}^2$ cross-sectional wire at a tip-wire separation of 5 nm, locally distorting the spin system and introducing dynamic topological defects as described above. The time dependent separation between the two 180\deg DWs formed by these defects was measured for a range of $q_{tip}$ values, with the results shown in figure \ref{sim} a). The tip first reaches the wire at $t = 2.5$ ns, soon after which the injected DWs are seen to rapidly separate. In the absence of magnetic wire-end charges the dipolar attraction between the two DWs eventually brings them back together, forming a stable composite 360\deg DW\cite{gartside2016novel}. The peak inter-DW separation reached was found to monotonically increase with $q_{tip}$. To investigate the mechanism by which this increasing separation occurs, a method is required for quantifying the amount of potential energy being stored in the contorted high-energy DW during the injection process. The key micromagnetic terms governing such dynamics are the exchange and demagnetisation energies. In order to separate the energy cost of the stable, non-distorted DWs from the additional potential energy driving the DW separation, the combined exchange and demagnetisation energies were measured from OOMMF simulations relative to the energy of a stable 360\deg DW where the spin-chains of both composite 180\deg DWs lie in a straight, low-energy conformation perpendicular to the long-axis of the wire. Any increase in micromagnetic energy relative to this can be linked to the distortions induced in the spin-chains by $H_{tip}$, with a value of zero reached at the end of the injection process corresponding to the formation of a stable 360\deg DW. The time-evolution of the resultant composite energy, termed the `excess micromagnetic energy', is plotted in fig. \ref{sim} b) for the same time period and $q_{tip}$ values as fig. \ref{sim} a) with a matching colour scheme indicating $q_{tip}$ for each curve. Time is presented on a logarithmic scale as the dynamics of interest occur in the initial stage of the injection process when the tip is close to the wire. Figure \ref{sim} b) shows an initial sharp increase in the excess micromagnetic energy as the vortex defect is introduced to the wire and the spin-chains are formed, peaking abruptly at $t = 3.2$ ns where the MFM tip finishes crossing the nanowire. At this point where the vortex defect decomposes and the DWs are no longer bound, the excess micromagnetic energy rapidly drops, coinciding with a large increase in the DW separation seen in fig. \ref{sim} a). This correlation serves as a good indication that the MFM tip is indeed inducing a large amount of micromagnetic potential energy in the nascent contorted DW, which is then converted to DW kinetic energy in the absence of the tip as the DW violently straightens out, generating momentum in the process. To better understand the relationship between the excess micromagnetic energy and the momentum imparted to the dynamic DW, figure \ref{sim} c) plots the peak DW separation reached in fig. \ref{sim} a) against the peak micromagnetic energy from fig. \ref{sim} b). The diamond markers represent data extracted from the OOMMF simulations shown in fig. \ref{sim} a) with colour codes corresponding to the same $q_{tip}$ values as in the previous two panels. A smooth monotonic increase in DW separation is observed with increasing excess micromagnetic energy, further suggesting that the release of pent-up micromagnetic potential is driving the DW separation. However, as micromagnetic dynamics are complex with multiple competing interactions present, it is not immediately clear that the excess micromagnetic energy is the dominant component driving the injection dynamics. In order to test whether the qualitative behaviour observed in the OOMMF simulations can be reproduced by considering just the excess micromagnetic energy, a simple semi-analytical model was employed, treating each 180\deg DW as a quasiparticle with effective mass $m = \frac{h^2 N}{4 \pi^2 K \lambda^2}$ where $N, K$ and $\lambda$ represent the number of spins in the DW, the transverse magnetic anisotropy energy of the wire and the DW width respectively\cite{doring1948tragheit,tatara2004theory,saitoh2004current}, and magnetic charge $2 q_m$ where $q_m = \pm M_s w t$ with $M_s$, $w$ and $t$ being the saturation magnetisation, wire width and thickness respectively\cite{hayward2010pinning}. Here, values of $m = 3.41 \times 10^{-23}$ kg and $2 q_m = 1.29\times 10^{-9}$ Am were determined. The two opposite magnetic charge polarity DWs are confined to a one-dimensional line representing the nanowire at an initial separation $r_0$, measured from the OOMMF simulations at the moment the vortex topological defect decomposes into half-vortex defects and the DWs become unbound. One DW remains fixed at $r = 0$, representing the low-energy DW formed from the straight spin-chain, while the other representing the high-energy DW formed from the contorted spin-chain is initialised moving away from $r = 0$ with a velocity $v_0$ given by converting the peak excess micromagnetic energy (taken from OOMMF at the point the tip-localised vortex defect decomposes) to kinetic energy via $\frac{1}{2} m v_0^2$. The moving DW feels an attractive magnetic Coulomb force $F_{DW} = \frac{\mu_0 }{4 \pi r^2}(2 q_m)^2$ directed towards the static DW at $r = 0$ and a viscous Gilbert-damping drag force\cite{kittel1956ferromagnetic} $F_{drag} = -vb$ where $b$ is a phenomenological damping coefficient, here $1.53 \times 10^{-14}~\mathrm{Nsm}^{-1}$. The equation of motion for the moving DW is then solved for DW position $r$ at each time step $\delta t = 1$ ps using numerical integration. The results of this model are plotted in figure \ref{sim} b) as the dotted grey line and can be seen to closely reproduce the behaviour observed in the OOMMF simulations. This provides further evidence supporting our hypothesis that micromagnetic potential energy stored in the shape-distortion of the forming DW by $H_{tip}$ is responsible for driving the momentum of the dynamic DW and therefore the writing process, rather like drawing back a pinball-machine plunger which is released once the tip completes its crossing of the wire. MTL injection on `quasi-infinite' nanowire arrays was performed with subsequent MFM imaging supporting the dynamics observed in OOMMF, suggesting that the picture of one high-energy DW `fired' down the wire and a second relatively static low-energy DW is indeed physically accurate. The results of this experiment are discussed below. 

To now examine the writing dynamics of MTL in a finite-length wire with end charges present, simulations were performed on $10~\mathrm{nm}~\times~75~\mathrm{nm}$ cross-sectional wires of 0.5-3 $\upmu$m length. As before, a magnetic charge and associated monopolar field representing the MFM tip were passed across the wire close to its central point. The topological defects and associated DWs were injected into the wire as described above, but for wires below a certain length threshold the dynamic DW now makes on contact with the wire end, unwinding to a smooth spin state before the remaining DW is pulled into the opposite end, also unwinding and leaving a collinear reverse-magnetised state. The longest reversible wire was determined for a range of $q_{tip}$ values by simulating injection on increasingly longer wires until the DWs were no longer able to reach the wire ends and reversal did not occur. The relation between $q_{tip}$ and the longest writeable wire is shown in fig. \ref{sim}d), with a monotonic increase observed in maximum writeable wire length with increasing $q_{tip}$ values. Wires several $\upmu$m in length are switchable by $q_{tip}$ values corresponding to typical high-moment MFM tips ($\sim 10^{-8} - 10^{-9}~\mathrm{Am}$), matching well with experimentally observed behaviour. The trend between longest writeable wire and $q_{tip}$ suggests that MTL may be tailored to switch wires of a desired length by selecting appropriate strength MFM tips, achieved practically by varying the thickness of the magnetic coating layer on the tip. Potentially allowing for longer wires to be written using thicker magnetic coatings ($>60$ nm) than are commercially available. 

\begin{figure}[tbp]
	\centering
	\includegraphics[width=14cm]{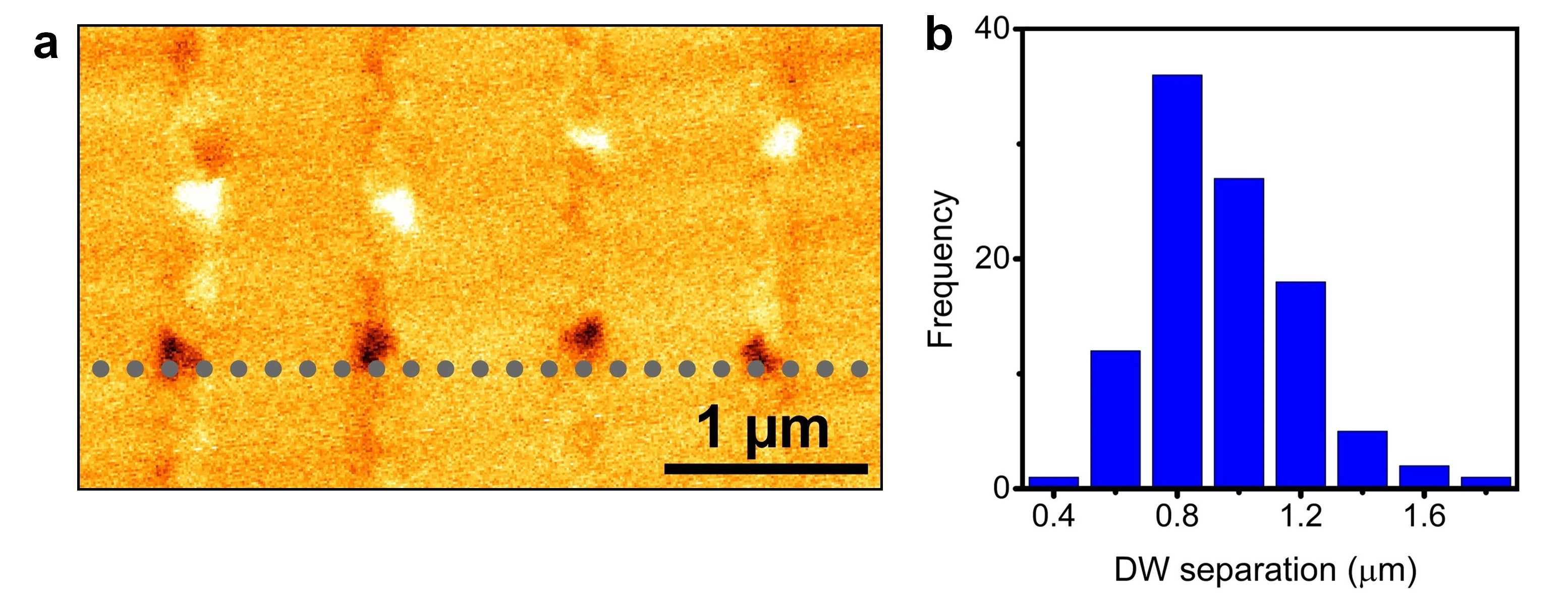}
\caption{\textbf{Results of MTL performed in `quasi-infinite' nanowire arrays}\\
\textbf{a)} MFM image of 100 nm $\times$ 10 nm cross-section Co nanowire array after MTL injection. Dotted grey line represents path of MFM tip during writing.\\
\textbf{b)} Histogram showing separation between injected DWs in nanowire arrays shown in a), measured over 101 MTL writing events.
}
\label{supp-hist}
\end{figure}

\section*{MTL in quasi-infinite nanowires}

To ascertain whether the injection dynamics observed in micromagnetic simulation represent a realistic picture of the injection process, 'quasi-infinite' nanowires were experimentally fabricated. Arrays of 20 $\upmu$m long nanowires were fabricated, such that at their central points the effects of the wire-end magnetic charges were negligible. The wires were initially magnetised along their long-axis by a global $H$ field. MTL injection was then performed across the mid-point of the wires and the resultant state imaged via MFM, with Co used rather than NiFe as the harder magnetic material allows for non-invasive MFM imaging of DWs without $H_{tip}$ disturbing them. A high deposition rate and coarse liftoff process were employed in an attempt to generate a high frequency of DW pinning sites in order to `trap' DWs at various stages of the dynamic injection process. Fig. \ref{supp-hist} a) shows a nanowire array after MTL injection. A 180\deg DWs is observed in each wire under the path of the tip, with a second 180\deg DW displaced 0.8-1.2 $\upmu$m along the wire. In each case the second DW is displaced to the same side of the DW lying under the tip path. The observed behaviour correlates well with 
our simulation results, suggesting that two DWs are created by the passing tip, one in a low-energy state which remains close to its initial location and one high-energy DW with a significant amount of momentum, travelling a distance away from its injected position. It is important to note that the DWs imaged here have become pinned at some point during the injection process (the lowest-energy state being a stable 360\deg DW with around 90 nm separation between DWs), each wire represents an effective snapshot at a random point during the injection process and the DW separations can be taken as a minimum separation reached. Figure \ref{supp-hist} b) is a histogram of DW separations, measured over 101 separate MTL injection events in the same nanowire shown in fig. \ref{supp-hist} a). A separation range of of 0.44-1.66 $\upmu$m is observed, corresponding well to the separations observed in OOMMF simulations and suggesting that dynamic topological defects induced by the tip are indeed responsible for the MTL writing process.

\begin{figure}[tbp]
	\centering
	\includegraphics[width=14cm]{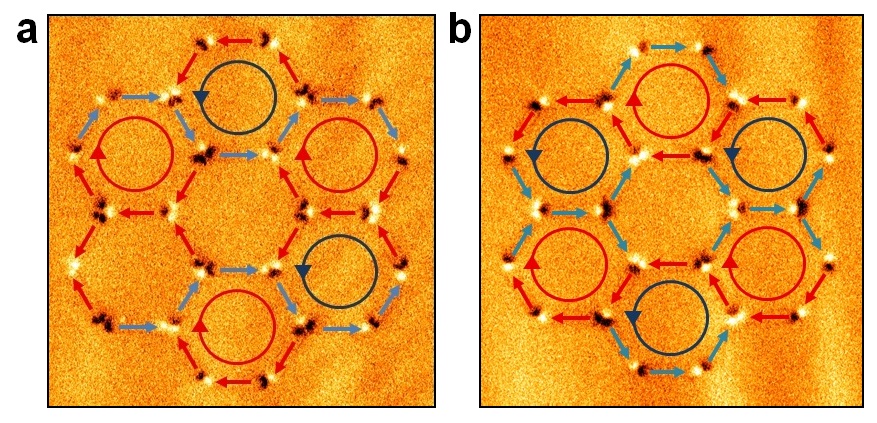}
\caption{\textbf{MTL-written states in disconnected ASI rosettes}\\
\textbf{a)} MFM image of a single bar (bottom-left loop) reversed from an otherwise complete ground state, highlighting `erase' functionality.\\
\textbf{b)} MFM image of alternate-chirality ground state. All 30 macrospins have been reversed from the original ground state.
}
\label{supp-mfm}
\end{figure}

\section*{Additional MTL-written states}

To demonstrate the `erase' functionality of MTL along with the ease by which complex states may be transitioned between, every macrospin in the spin-crystal ground state shown in fig. 3 b) was reversed via MTL, preparing the alternate chirality ground state. Figure \ref{supp-mfm} a) shows a single lithographically reversed bar in an otherwise complete ground state configuration. The remaining 29 macrospins were then reversed using MTL. Figure \ref{supp-mfm} b) showing the resultant alternate-chirality ground state, highlighting the accuracy and flexibility of the writing technique. 

\section*{Micromagnetic simulation videos}

Micromagnetic simulation videos of various MTL processes are available in the following YouTube playlist:

https://www.youtube.com/playlist?list=PLtNlYiPezEZ9i69fsTJPjOr5snyaLJZMt

\textbf{Supplementary video 1} MFM tip of $q_{tip} = 10^{-9}$ Am crosses a right-magnetised 1~$\mu$m $\times~75~\times~10~\mathrm{nm}^2$ nanowire, resulting in a high-energy DW accelerating left and acheiving magnetisation reversal.

\textbf{Supplementary video 2} MFM tip of $q_{tip} = 10^{-9}$ Am crosses a left-magnetised 1~$\mu$m $\times~75~\times~10~\mathrm{nm}^2$ nanowire, resulting in a high-energy DW accelerating right and acheiving magnetisation reversal.

\textbf{Supplementary video 3} MFM tip of $q_{tip} = -10^{-9}$ Am crosses a right-magnetised 1~$\mu$m $\times~75~\times~10~\mathrm{nm}^2$ nanowire, resulting in a high-energy DW accelerating right and acheiving magnetisation reversal.

\textbf{Supplementary video 4} MFM tip of $q_{tip} = -10^{-9}$ Am crosses a left-magnetised 1~$\mu$m $\times~75~\times~10~\mathrm{nm}^2$ nanowire, resulting in a high-energy DW accelerating left and acheiving magnetisation reversal.

\textbf{Supplementary video 5} MFM tip of $q_{tip} = 1.9 \times 10^{-9}$ Am crosses a left-magnetised 1~$\mu$m $\times~75~\times~10~\mathrm{nm}^2$ nanowire, resulting in a high-energy DW rapidly accelerating right and acheiving magnetisation reversal.

\textbf{Supplementary video 6} MFM tip of $q_{tip} = 10^{-9}$ Am crosses a left-magnetised 1.3~$\mu$m $\times~75~\times~10~\mathrm{nm}^2$ nanowire, resulting in a high-energy DW accelerating right. The DW lacks the energy to achieve magnetisation reversal, instead forming a 360\deg DW which is pulled to the right of the wire before annihilating.

\section*{Methods}

\subsection*{Structures and fabrication}

\begin{figure}[tbp]
	\centering
	\includegraphics[width=14cm]{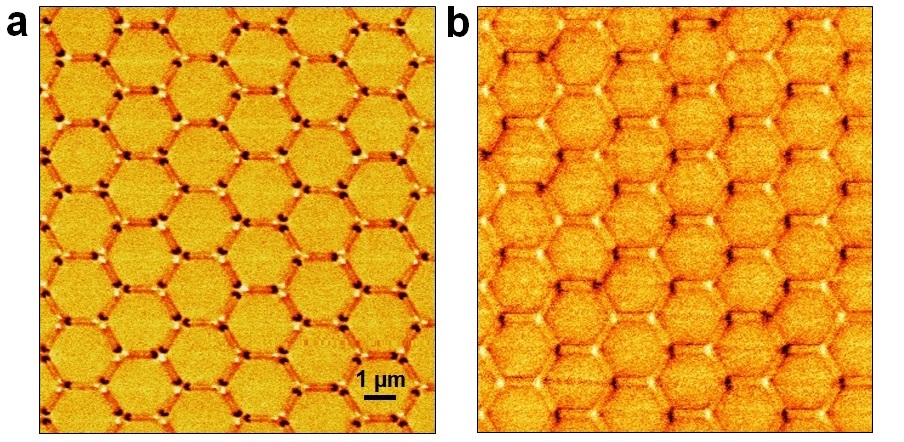}
\caption{\textbf{MFM images of as-grown ASI arrays.}\\
\textbf{a)} Disconnected and \textbf{b)} connected ASI arrays immediately after deposition. The ice-rules are obeyed at every vertex indicating a strongly interacting system.
}
\label{supp-demag}
\end{figure}

Nanostructures were fabricated using an electron beam lithography lift-off process with PMMA resist. Permalloy (nominally Ni$_{81}$Fe$_{19}$) of 10 nm thickness was deposited by thermal evaporation onto Si/SiO$_2$ substrates. 

To confirm that ASI arrays were in the strongly-interacting regime, structures were imaged via MFM immediately after deposition. In a randomly prepared non-interacting ASI system where the orientation of each macrospin is independent of its neighbours, the chance of observing no $\rpm 3~q_m$ monopole-defect vertices is given by the ratio of ice-rule obeying microstates $\frac{3}{\sqrt{2 }}^N$, where $N$ is the number of vertices, to the total number of microstates $2^\text{n}$ where $n$ is the number of macrospins. A non-interacting 15 $\upmu$m x 15 $\upmu$m array containing 118 vertices and 222 macrospins therefore has a $2.1^{-11}$ probability of containing no $\pm 3~q_m$ vertices. Figure \ref{supp-demag} shows MFM images of disconnected (a) and connected (b) as-grown arrays with no ice-rule violations observed, showing structures are in the strongly-interacting ASI regime.

\subsection*{Magnetic force microscopy}

The magnetic charge states of nanostructures were written and measured using MFM with HM and LM tips respectively. MFM is directly sensitive to magnetic charge \cite{Hubert-MFM-charge}, providing an ideal tool to read and write magnetic charge landscapes. Two MFM systems were used, a Dimension 3100 and an Asylum MFP-3D. An interface was developed to define the path of the tip relative to the nanopatterned structures (and therefore the magnetic state to be written) graphically, by drawing over an existing AFM/MFM image or CAD schematic, or by defining an $(x,y)$ position list. Writing does not occur when the tip-sample separation is over 5-10 nm, allowing for free movement of the tip around samples between writing events by raising the tip. Writing operations were performed with the tip moving perpendicular to the wire length. MFM imaging was performed in a tapping mode at lift-heights of 35-45 nm. The HM and LM tips have moments and stray fields of {\raise.17ex\hbox{$\scriptstyle\mathtt{\sim}$}}$5\times{10}^{-13}$ emu, 690 Oe and {\raise.17ex\hbox{$\scriptstyle\mathtt{\sim}$}}$3\times{10}^{-14}$ emu, 320 Oe respectively\cite{jaafar2008calibration}, with stray fields measured at a typical AFM tip-sample separation ({\raise.17ex\hbox{$\scriptstyle\mathtt{\sim}$}}2-5 nm) from the tip apex. Experimental determination of the monopole-like magnetic charge associated with MFM tips is challenging, but prior work using Lorentz electron tomography have arrived at values of {\raise.17ex\hbox{$\scriptstyle\mathtt{\sim}$}}${10}^{-8}$ Am\cite{mcvitie2001quantitative}, similar to those simulated here.
Writing has been succesfully performed on wire-like structures of widths and lengths up to 150 nm and 4 $\upmu$m with tip speeds of up to 400 $\upmu\textrm{m/s}$. Significantly faster tip speeds of up to 200 $\textrm{m/s}$ show succesful writing in micromagnetic simulations but cannot yet be experimentally tested using our current equipment.

\subsection*{Micromagnetic simulation}

Additional insight into the reversal dynamics of the nanowires was provided by performing a series of micromagnetic simulations using the object-oriented micromagnetic framework (OOMMF)\cite{oommf}. Typical micromagnetic parameters for permalloy were used, i.e. saturation magnetisation, $M_S = 860 \times 10^{3}$~A/m, exchange stiffness, $A = 13 \times 10^{-12}$~J/m, zero magnetocrystalline anisotropy and a Gilbert damping parameter, $\alpha = 0.01$. The point probe approximation (that at small tip-sample separations an MFM tip may be described by a point monopole moment\cite{Phys.Lett.A_137_475, JAP_86_3410}) was used. This approximation is widely used in MFM simulations and previous work analysing systems with similar dynamics has shown that both dipolar and monopolar-modelled tip fields induce tip-localised magnetisation vortices in ferromagnetic thin films \cite{magiera2012magnetic,magiera2014magnetic}.

The simulated nanowires were 10~nm thick with widths of 75~nm and were divided into $2.5 \times 2.5 \times 10$~nm cells. Finite-length wires of 0.5-3~$\mu$m length were studied along with a semi-infinite wire of 10~$\mu$m length where the demagnetisation effects from the wire ends were corrected for by the inclusion of plates of fixed magnetic charge at the nanowire ends\cite{IEEE.Trans.Magn_33_4167}.

The field from the MFM tip was modelled as a single magnetic charge, $q_{tip}$, producing a radial field $H = \frac{\mu_0}{4\pi} \frac{q_{tip}}{r^2}$ at a distance $r$ from the charge. During the simulation this magnetic charge moved perpendicular to the nanowire axis in 1~nm steps every 10~ps representing a velocity of 100~m/s.  This is faster than the velocities of $\sim 10^{-4}~m/s$ investigated experimentally. However, the simulated speed is well below those associated with exciting any precessional spin modes and is believed to be reasonable in this case. To avoid a discrete jump in applied field on the nanowire when starting the simulation, the magnetic charge was initialised 300~nm away from the nanowire in the plane of the wire. Following the magnetic charge interaction with the wire, the wire was allowed to relax to an energetically stable state in zero field to obtain its final configuration.

\bibliography{library}